\documentstyle[preprint,prc,aps,epsfig]{revtex}

\tighten

\begin{document}
\title{Shell model description of ``mixed-symmetry'' states in $^{94}$Mo}
\author{A.~F.~Lisetskiy$\,^1$, N. Pietralla$\,^{1,2}$, 
 C.~Fransen$\,^1$, R.~V.~Jolos$\,^{1,3}$, P. von Brentano$\,^1$}

\address{$^1\,$ Institut f\"ur Kernphysik, Universit\"at zu K\"oln, 
                50937 K\"oln, Germany \\
         $^2\,$ Wright Nuclear Structure Laboratory, Yale University, 
                New Haven, CT 06520-8124\\ 
         $^3\,$ Bogoliubov Theoretical Laboratory, 
        Joint Institute for Nuclear Research, \\
        141980 Dubna, Russia}

\date{\today}
\maketitle

\begin{abstract}
Shell model calculations have been performed for the nucleus $^{94}$Mo. 
The calculated excitation energies and electromagnetic properties of 
low-lying states are in good agreement with the data, which include 
states with mixed-symmetry (MS) assignments in previous 
Interacting Boson Model studies. 
In the shell model large isoscalar $E2$ matrix elements are found between 
states with MS assignments indicating that they form a class of 
states with similar proton-neutron symmetry. 
\end{abstract}

\pacs{21.60.Cs, 21.10.Re, 23.20.-g, 27.60.+j}

{\bf Keywords:} Shell model, $^{94}$Mo, mixed symmetry states.

 
\section{Introduction}

Recently performed photon scattering experiments and  
$\gamma\gamma$--coincidence studies \cite{Pietr,PiFr3+} of 
the nucleus $^{94}$Mo indicate the 
existence of low-lying valence shell excitations with 
proton-neutron symmetry different to that of the ground state 
in that nucleus. 
The measurements \cite{Pietr,PiFr3+} of absolute $E2$ and $M1$ 
transition strengths have been interpreted in terms of 
$J^\pi = 1^+, 2^+, 3^+$ mixed-symmetry states in the 
framework of the proton-neutron version \cite{Arima} of 
the Interacting Boson Model (IBM-2). 
The proton-neutron symmetry of an IBM-2 wavefunction is quantified 
by the $F$-spin quantum number \cite{Arima,OtAr78}, which is the isospin 
for basic proton and neutron bosons. 
The IBM-2 predicts the lowest-lying collective states 
to be dominantly isoscalar excitations of the almost 
proton-neutron symmetric ground state with maximum $F$-spin 
quantum number $F = F_{\rm max}$. 
This is put in evidence by the existence of $F$-spin multiplets 
with rather constant energies \cite{BrGe85}. 
The IBM-2 predicts also valence shell excitations with wavefunctions, which 
are not symmetric with respect to the proton-neutron degree of 
freedom \cite{Arima,OtAr78,TakaPhD,Iach84}. 
Such states have $F$-spin quantum numbers $F<F_{\rm max}$ 
and are called {\em mixed-symmetry} (MS) states. 
The observations for the $1^+_1, 2^+_3, 3^+_2$ states of $^{94}$Mo 
agree with MS assignments \cite{Pietr,PiFr3+}. 
The key-signatures used for the assignments of MS character to 
these states of  $^{94}$Mo were the measured relatively strong 
$M1$ transitions and weakly-collective $E2$ transitions 
to low-lying symmetric states.

Separate proton and neutron quadrupole surface vibrations, which lead to 
eigenstates with different symmetries with respect to proton-neutron 
permutations, have been considered in a geometrical model already in 
the sixties by Faessler \cite{Faess66}. 
At that time these states were predicted to exist above the particle 
threshold, at a much higher energy than recently observed. 
In the eighties \cite{Faess86,Faess87} the isovector vibrational model 
was improved giving quantitatively better description of the $2^+$ isovector 
vibrational states.  
The existence of collective orbital isovector $M1$ transitions in the 
geometrical Two Rotor Model for deformed nuclei was realized by Lo Iudice 
and Palumbo \cite{LoIudice}. 
A more general type of enhanced magnetic dipole transitions in the 
valence shell of all open--shell nuclei, not only deformed, has 
been predicted \cite{Arima,OtAr78,TakaPhD,Iach84} in the IBM-2 as 
the decays of MS states. 
Following these predictions the dominantly isovector $J^\pi=1^+$ 
state, generally called scissors mode, was discovered by Richter 
{\em et al.} \cite{Bohle84} in inelastic electron scattering 
experiments in Darmstadt. 
The typically fragmented $1^+_{\rm sc}$ scissors mode was further 
investigated mainly by electron scattering \cite{Richter}, 
photon scattering \cite{KnPi96} and neutron scattering \cite{Yates} 
experiments.

Many theoretical studies were published to explain  
the structure of this mode, e.g. 
\cite{Lippa,Bes,Dra,Faessler,Sugawa,Hama,Fill}. 
The $1^+$ mode is expected to be dominantly excited by the isovector 
part of the $M1$ operator indicating its isovector character. 
The large $M1$ transition strength and its close correlation 
\cite{Rang91,PiM1E2,pvnc,PieEsc} to the collective $E2$ excitation strength 
in deformed nuclei is usually considered an indication of the collective 
nature of the $1^+_{\rm sc}$ state. 
Another state with spin different from $J^\pi=1^+$ but of 
similar isovector character, the one-quadrupole phonon $2^+_{\rm ms}$ 
state, has been identified from $M1$ strengths, too. 
The first $2^+_{\rm ms}$ assignments to states at about 2 MeV in vibrational 
nuclei around the $N=82$ shell closure were based only on small $E2/M1$ 
multipole mixing ratios \cite{Hamilton,Park,Molnar,Gianna}
Lateron, several of these assignments were confirmed by the measurement 
of relatively large $M1$ and weakly-collective $E2$ transition 
strengths \cite{Vermeer,Fazekas,Vanhoy,PiBa136}. 
Based on measured $M1$ and $E2$ strengths the experimentally observed 
$1^+_1$, $2^+_3$, and $3^+_2$ states in the nucleus $^{94}$Mo were 
argued to belong to this type of excitations, too \cite{Pietr,PiFr3+}.

In nuclei not too far from shell closures the structure of 
low-lying valence excitations including the states, which are outside of 
the IBM configurational space, can be described using the 
shell model.  
It is the main aim of the present paper to describe  the 
structure of the states observed in \cite{Pietr,PiFr3+} in the framework of 
the nuclear shell model, especially the structure of the $1^+_1$, 
$2^+_3$ and $3^+_2$ states. 
We present the results of shell model calculations for 
the low-lying positive parity states of $^{94}$Mo with 
spin quantum numbers $J^\pi = 0^+$ -- $4^+$ and 
we discuss the isotensor character of their electromagnetic 
transitions.

\section{Theoretical Approach}
The shell model Hamiltonian is taken as
$H=H_0+V$
where the mean field is given by
\begin{equation}
H_0=\sum_{k}^{n_{\rm val}}\varepsilon_k a^+_{\rho_k} a_{\rho_k},
\end{equation}
where $n_{\rm val}$ is a number of single particle states in the adopted 
valence shells and the residual interaction is 
\begin{equation}
V=\sum_{\rho_a, \rho_b,\rho_c,\rho_d,J,M,T} <(\rho_a \rho_b)_{JT}|V_{12}|(\rho_c \rho_d)_{JT}> 
(a^+_{\rho_a}a^+_{\rho_b})_{JM}^T (a_{\rho_c}a_{\rho_d})_{JM}^T.
\end{equation}
Here $a^+_{\rho}$ creates and $a_{\rho}$ annihilates a particle in the
single particle orbital $|\rho\rangle \equiv |n,l,j,m_j,t=1/2,t_z\rangle$ and 
T is the isospin of the coupled particles.
The first term $H_0$ is the Hamiltonian of
the noninteracting particles. 
The residual interaction we used is the Surface Delta Interaction 
(SDI) \cite{Mos}. 
This interaction contains strong pairing and quadrupole parts and 
higher multipolarity components, which are weaker than the first two. 
The SDI is an extremely simple interaction from the mathematical 
point of view. 
The two body matrix elements of the SDI are:
\begin{equation}
\label{sdi} 
<\rho_a \rho_b|V_{SDI}({\bf r}_1,{\bf r}_2)|\rho_c \rho_d>_{JT}=-4\pi A'_T 
<\rho_a \rho_b|\delta(\Omega_{12})\delta(r_1-R)\delta(r_2-R)|\rho_c \rho_d>_{JT}
\end{equation}
where $\Omega_{12}$ is the angle between the interacting particles, 
$R = 1.2\,A^{1/3}$ fm is the nuclear radius, and $A'_T$ is the 
strength constant of the SDI.
There are three parameters
${A'}_{T=1}^{\rho',\rho}$ ($\rho,\rho' \in \{p,n\}$) that describe 
the interaction in the T=1 channel and one parameter ${A'}_{T=0}^{pn}$ 
describing the interaction in the T=0 channel.
The fitted interaction 
parameters $A_{T}^{\rho',\rho}$ are
connected to the parameters from Eq.(\protect\ref{sdi}) by the following
expression:$A_{T}^{\rho',\rho}$=
$A_{T}'^{\rho',\rho}<\delta(r_\rho-R)\delta(r_{\rho'}-R)>$, 
where the radial matrix
elements $<\delta(r_\rho-R)\delta(r_{\rho'}-R)>$ are supposed to be
independent of the single particle states involved (see for details
\cite{bru77}).

For the shell model description of $^{94}$Mo one may want  
to consider the $N=Z=50$ closed shell nucleus $^{100}$Sn 
as the inert core. 
In order not to have to treat a too large model space we consider 
eight proton holes in the proton shells $\pi g_{9/2}$ and 
$\pi p_{1/2}$ for the description of $^{94}_{42}$Mo. 
Because of the Pauli principle this problem is equivalent to 
the consideration of four proton particles in these shells 
outside of the core $^{88}_{38}$Sr$_{50}$. 
Therefore, we adopt $^{88}$Sr as the inert core for the 
following shell model description of $^{94}$Mo.

\begin{figure}[bt]
\epsfxsize 15.5cm \centerline{\epsfbox{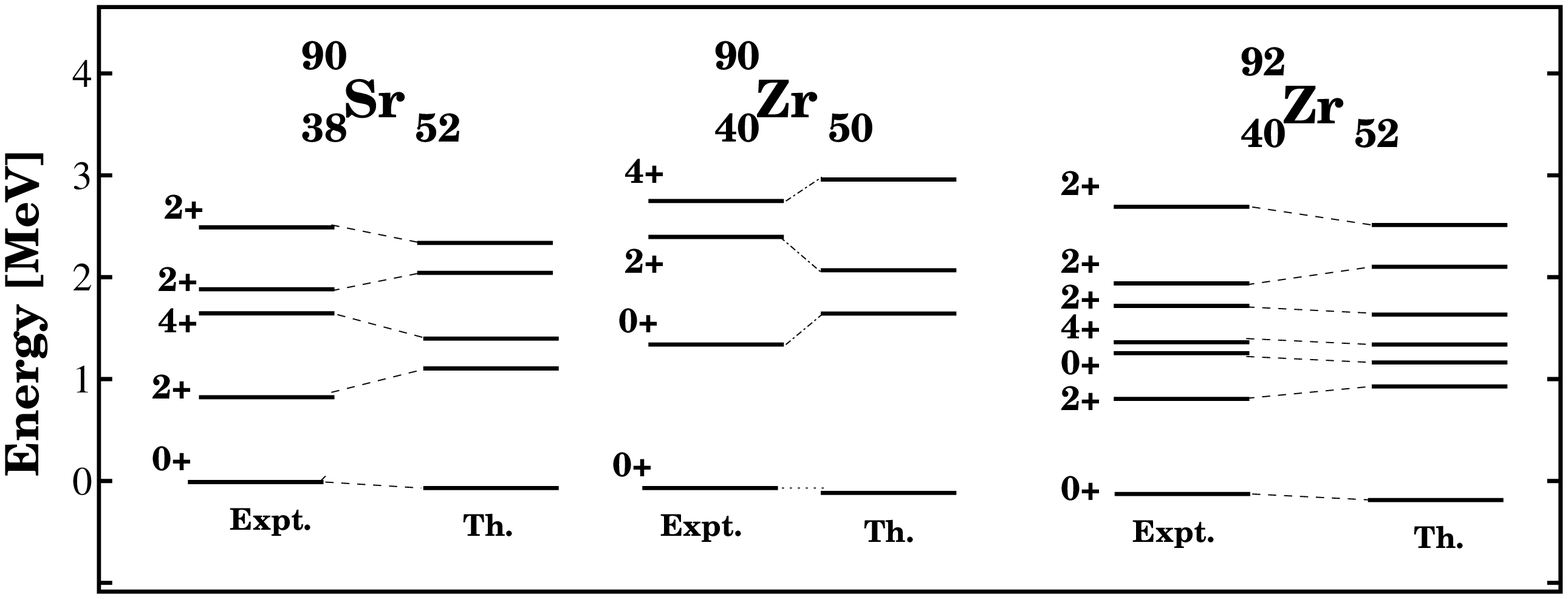}}
\caption{Calculated and experimental positive parity 
\protect$J^\pi = 0^+$--\protect$4^+$ states below 3 MeV in $^{90}$Sr, 
 $^{90}$Zr, and $^{92}$Zr.}
\label{90Sr} 
\end{figure}

Single-particle energies $\varepsilon_j$ were obtained from 
calculations for the neutron--odd nuclei $^{89}$Sr, $^{91}$Zr 
and $^{93}$Mo and for the proton--odd nuclei 
$^{89}$Y, $^{91}$Y, $^{91}$Nb, $^{93}$Nb, and $^{93}$Tc. These 
single-particle energies are close to those from \cite{Lederer}.
In order to get a rough estimate of the values of 
$A_{T=1}^{nn}$ and $A_{T=1}^{pp}$ parameters 
shell model calculations have been 
performed for the isobars $^{90}$Sr and $^{90}$Zr, which have either 2 
neutrons or 2 protons, respectively, outside the core $^{88}$Sr 
(see Fig.\ref{90Sr}). 
The results are $A_{T=1}^{nn}=0.23$ MeV and $A_{T=1}^{pp}=0.35$ MeV. 
The $A_{T=1}^{pn}$ parameter is taken as the 
average of  $A_{T=1}^{nn}$ and $A_{T=1}^{pp}$, i.e. 
$A_{T=1}^{pn}=(A_{T=1}^{nn}+A_{T=1}^{pp})/2$. 
The value of 0.48 MeV for the $A_{T=0}^{pn}$ parameter was 
obtained by a fit to the excitation energies of the nucleus $^{92}$Zr, 
which contains 2 protons and 2 neutrons outside the core $^{88}$Sr.  
Experimental and calculated low-spin level schemes for the three 
even-even nuclei $^{90}$Sr, $^{90}$Zr, and $^{92}$Zr are shown 
in Fig.\ref{90Sr}.  
The final values of the SDI parameters used for the description 
of $^{94}$Mo were optimized by a fit to the low-spin level scheme 
of $^{94}$Mo and are close to the values above. It is interesting to note 
that the excitation energies of low-lying states are much less sensitive to 
the $A_{T=1}^{pn}$ parameter than to the $A_{T=0}^{pn}$ parameter. However 
the electromagnetic transition strengths are very sensitive to both 
$A_{T=1}^{pn}$ and $A_{T=0}^{pn}$ parameters and the above discussed choice 
gives the best agreement between calculated and experimental strengths in 
$^{94}$Mo nucleus.
The single-particle energies and SDI parameters used for $^{94}$Mo 
are presented in Table \ref{parameters}. 
The SM calculations were performed on the Cologne Sun Ultra 
Enterprise 4000 workstation with two 166 MHz Ultra Sparc 
processors using the code RITSSCHIL \cite{Zwartz}.

\section{Discussion}

\begin{figure}[bt]
\epsfxsize 13.5cm \centerline{\epsfbox{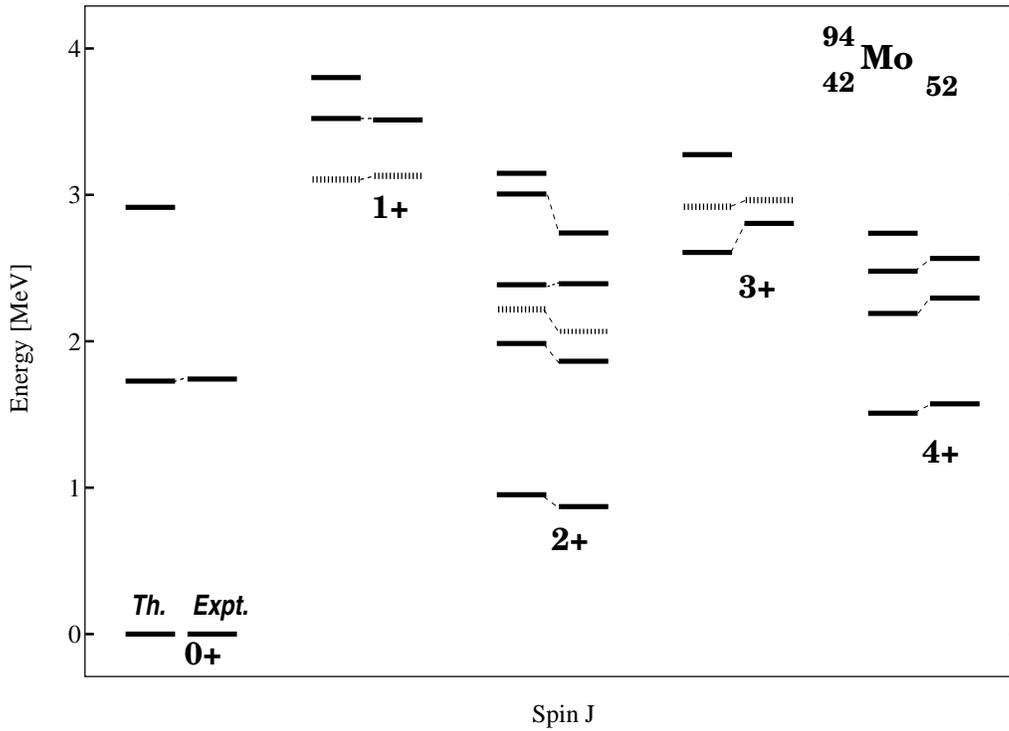}}
\caption{Comparison of calculated and experimental spectra of the
         \protect$J^\pi = 0^+$--\protect$4^+$ states in $^{94}$Mo. 
         The states with MS assignments are plotted with dashed lines.}
\label{fig:LevMo94} 
\end{figure}

In the present shell model calculation for $^{94}$Mo two neutrons 
can be distributed among five single particle orbitals: 
$2d_{5/2}$, $3s_{1/2}$, $1g_{7/2}$, $2d_{3/2}$ and $1h_{11/2}$. 
The lowest neutron orbital is $2d_{5/2}$. 
The fully occupied neutron orbital $1g_{9/2}$ forms the 
$N=50$ closed shell 
and is about 4 MeV below the neutron $2d_{5/2}$ orbital.  
Therefore, the influence of the $1g_{9/2}$  orbital on the 
structure of low--lying 
states of $^{94}$Mo is expected to be small \footnote{This expectation is 
supported by the first results of large scale shell model calculations 
for \protect$^{94}$Mo assuming the $Z=N=40$ nucleus \protect$^{80}$Zr as 
the inert core, which yield at the present stage almost negligible 
contributions of the \protect$\nu(g_{9/2})$ orbital to low-lying low-spin 
states of \protect$^{94}$Mo \protect\cite{EC00pc}.}. 
This justifies the assumption of the $N=50$ neutron core. 
For protons we have included the two orbitals $1g_{9/2}$ and 
$2p_{1/2}$ in the configurational space. 
The closest higher--lying proton orbital to the proton $1g_{9/2}$ orbital 
appears more then 4 MeV above the Z=50 closed shell and is neglected.  
But the proton $p_{1/2}$ orbital is much closer to the proton $1g_{9/2}$ -- 
about 1 MeV lower -- and we have taken it into account by choosing 
$Z=38$ as the proton core. 
Within this configurational space we reproduce well many of the 
excited states in the spectrum of $^{94}$Mo. 
This is shown in Fig. \ref{fig:LevMo94}.

The main components of the low-lying states (see Table \ref{structure})  
are seniority 2 and 4 with two protons in $\pi(1g_{9/2})$ and two neutrons 
in  $\nu(2d_{5/2})$. 
But the influence of the $\pi(2p_{1/2})$ proton orbital 
is also significant - almost all states contain large components 
$\pi (p_{1/2}^{-2}g_{9/2}^4)_J$ and this is the main component 
for the $0^+_2$ state. 
The contributions of the neutron orbitals  $\nu(3s_{1/2})$, $\nu(1g_{7/2})$, 
$\nu(2d_{3/2})$ and $\nu(1h_{11/2})$ are smaller but for the 
$1^+_2$ the component $\nu(g_{7/2}^1d_{5/2}^1)$ is already the main one.

The $M1$ and $E2$ transition probabilities have been calculated and 
compared with the new experimental data. 
The results are shown in Tables \ref{transm1} and \ref{transe2}. 
The reproduction of the data is in most cases very good. 
Tables \ref{transm1} and \ref{transe2} include also IBM-2 predictions 
in the  O(6) dynamical symmetry limit, where the states 
$1^+_1, 2^+_3, 3^+_2$ have mixed-symmetry. We note that the IBM-2 has 
only one free parameter - the effective quadrupole proton boson charge $e_p$,
while the corresponding neutron charge was put to zero $e_n=0$.  
It is remarkable how well the shell model agrees also with the 
IBM-2 with only a few exceptions. The agreement of shell model 
calculations with IBM-2 results in different mass regions was noted 
also by other authors (see, for instance, \cite{Scho85,Liu87}).

Let us discuss now the calculated $M1$ and $E2$ transition strengths 
in more detail. 
The MS assignments for the $1^+_1, 2^+_3, 3^+_2$ states of $^{94}$Mo 
were based on the measurements of relatively large $M1$ transition 
strengths. 
For the calculation of $M1$ transitions between the shell model states
 we consider a nuclear magnetic dipole operator, which is 
the sum of proton and neutron one-body terms for orbital and spin 
contributions:
\begin{equation} 
\label{eq:TM1} 
{\bf T}(M1)= \sqrt{3 \over 4 \pi } 
     \left( \sum_{i=1}^{Z}\left[g_p^l{\bf l}^p_i+g_p^s{\bf s}^p_i \right]+
            \sum_{i=1}^{N}\left[g_n^l{\bf l}^n_i+g_n^s{\bf s}^n_i\right] 
     \right)\ \mu_N, 
\end{equation}
where $g_{\rho}^l$ and $g_{\rho}^s$  are the orbital and spin g-factors 
and ${\bf l}^\rho_i$, ${\bf s}^\rho_i$ are the single particle orbital angular 
momentum operators and spin operators. 
For further discussion it is useful to decompose the $M1$ operator 
into an isoscalar part 
\begin{equation}
\label{eq:TM1IS} 
{\bf T}_{IS}(M1)=\sqrt{3 \over 4 \pi }  
     \left( g_J {\bf J}+ g_S {\bf S} \right) \ \mu_N, 
\end{equation}
and an isovector part 
\begin{eqnarray}
\label{eq:TM1IV} 
{\bf T}_{IV}(M1)=\sqrt{3 \over 4 \pi } 
     \left( {g_p^l - g_n^l\over 2} \left[{\bf L}_p - {\bf L}_n \right]+
  {g_p^s - g_n^s \over 2}\left[{\bf S}_p - {\bf S}_n \right] \right). 
\end{eqnarray} 
where  ${\bf L}_\rho$ and  ${\bf S}_\rho$ denote the total orbital 
angular momentum and total spin operators for protons 
($\rho=p$) and neutrons ($\rho=n$). 
${\bf J}={\bf L} + {\bf S}$ is the total angular momentum and does 
not generate $M1$ transitions. 
$g_J = (g_p^l + g_n^l)/2 = 1/2$ and 
$g_S = [g_p^s + g_n^s -(g_p^l + g_n^l)]/2 = 0.88\alpha_q -1/2$ with the 
quenching factor $\alpha_q$ defined by $g_\rho^s = \alpha_q g_\rho^{s,{\rm free}}$. 
The free spin $g$-factors are $g_p^{s,{\rm free}} = 5.58$ and 
$g_n^{s,{\rm free}} = - 3.82$. 
Since $g_p^s$ and $g_n^s$ are of opposite sign and comparable, 
the isoscalar nondiagonal piece of an $M1$ matrix element is 
usually very small. 
It vanishes exactly for a quenching factor $\alpha_q = 0.57$. 
For a good reproduction of the measured $M1$ transition strengths 
(see Table \ref{transm1}) and for the sake of easy interpretation we used 
the quenching factor $\alpha_q = 0.57$ which results in pure isovector $M1$ 
transitions.

The isovector $M1$ ground state excitation strength is calculated to be 
concentrated in the $1^+_1$ state. 
This agrees with the identification of the $1^+_1$ state with the 
scissors mode in the nucleus $^{94}$Mo. 
This identification receives further support from the consistency 
\cite{Pietr} of the data on $^{94}$Mo with the systematics of the 
scissors mode extrapolated from the deformed rare earth nuclei. 
However, one cannot ascribe the collective scissor mode in the 
near-spherical nucleus $^{94}$Mo as a pure orbital mode. 
Our calculations show that spin and orbital contributions are 
almost equal which agrees with the results of a single 
$j$ shell model \cite{Zamick85}.

The key signature for the lowest $2^+_{\rm ms}$ state, with a 
proton-neutron symmetry similar to that of the scissors mode, 
is a strong $M1$ transition to the $2^+_1$ state. 
Therefore, it is interesting to look to the $2^+_1 \to 2^+_i$ 
$M1$ strength distribution to judge the possible fragmentation  
of the $2^+_{\rm ms}$ state. 
>From the data it follows that the $2^+_3\rightarrow 2^+_1$ is the 
strongest $M1$ transition 
from an excited $2^+$ state to the $2^+_1$ state which led to the MS 
assignment for the $2^+_3$ state. 
Comparing the calculated $M1$ strengths of the 
$2^+_2\rightarrow 2^+_1$, $2^+_3\rightarrow 2^+_1$ and 
$2^+_4\rightarrow 2^+_1$ transitions one notes that the 
calculated $B(M1;2^+_3\rightarrow 2^+_1$) value is about 
five times larger than the other two. 
The dominance of the $B(M1;2^+_3\rightarrow 2^+_1$) value agrees 
with the data. 
The calculated $B(M1;2^+_3\rightarrow 2^+_1$) value is also more 
than four times larger than the calculated 
$B(M1;2^+_5\rightarrow 2^+_1$) value, which already overestimates the data. 
The shell model calculation agrees with the observation that 
the $2^+_3 \to 2^+_1$ transition concentrates the $M1$ strength 
between excited $2^+$ states and the $2^+_1$ state. 
This  relatively strong isovector $M1$ transition agrees with 
the MS assignment for the $2^+_3$ state.

Also the experimental $3^+_2$ state decays by relatively strong $M1$ transitions 
to low-lying states. 
The calculated excitation energy of the $3^+_2$ state matches the experimental 
energy within 50 keV whereas the $3^+_1$ state lies 120 keV lower and the $3^+_3$ 
state lies 350 keV higher than the experimental $3^+_{\rm ms}$ state. 
Therefore we compare the $3^+_2$ shell model state with the observed 
$3^+_2$ state at 2965 keV. 
We note that the measured strong $M1$ transition from the $3^+_2$ state to 
the $4^+_1$ state is reproduced by the shell model within the experimental 
error bar. 
The shell model, however, underestimates the $M1$ strength of the 
$3^+_2 \to 2^+_2$ transition by a factor of two, while it overestimates the 
$3^+_2 \to 2^+_1$ $M1$ transition strength by an order of magnitude. 
The shell model results for the $3^+_2$ state disagree not only somewhat with 
the data but also with the prediction of the IBM in the O(6) dynamical 
symmetry limit for the $3^+_{\rm ms}$ state. 
However, the $3^+_1$ and $3^+_2$ states are close in energy, which renders 
the calculation of the wave functions more uncertain in the shell model, where 
no quantum number like the $F$-spin exists, which can assure the 
orthogonality of MS states to symmetric states. 
Moreover, the calculated $3^+_3$ state also shows $M1$ and $E2$ 
properties which are close to those of the experimental $3^+_2$ state. 
We conclude that the $3^+_{\rm ms}$ character is spread about the 
first three $3^+$ states in our shell model calculation with the 
surface delta residual interaction and the $^{88}$Sr core. 
Thus for the $3^+_2$ state the MS assignment from the shell model results 
is less clear.
We consider this fact not as an argument against the MS assignment of 
the experimental $3^+_2$ state of $^{94}$Mo but as an indication 
of the limit of our present shell model approach for the description 
of the details (and the mixing) of wave functions for states, which 
lie close in energy. 
In total the $M1$ transition strengths calculated in the shell model 
support (or at least do not disagree with) the MS assignments for the 
$1^+_1$, $2^+_3$, and $3^+_2$ states of $^{94}$Mo, if the existence of 
relatively strong isovector $M1$ transition is considered as a 
sufficient argument in favor of MS structures.

However, it should be stressed that the strongest M1 transition found 
in $^{94}$Mo nucleus \cite{Fr4+} connects the $4^+_2$ and $4^+_1$ states 
that is in agreement with the present shell model calculations. 
This transition falls out from the $sd$-IBM-2 scheme and probably is 
related to the excitations of $g$-bosons in terms of the IBM.

\begin{figure}[bt]
\epsfxsize 12.5cm \centerline{\epsfbox{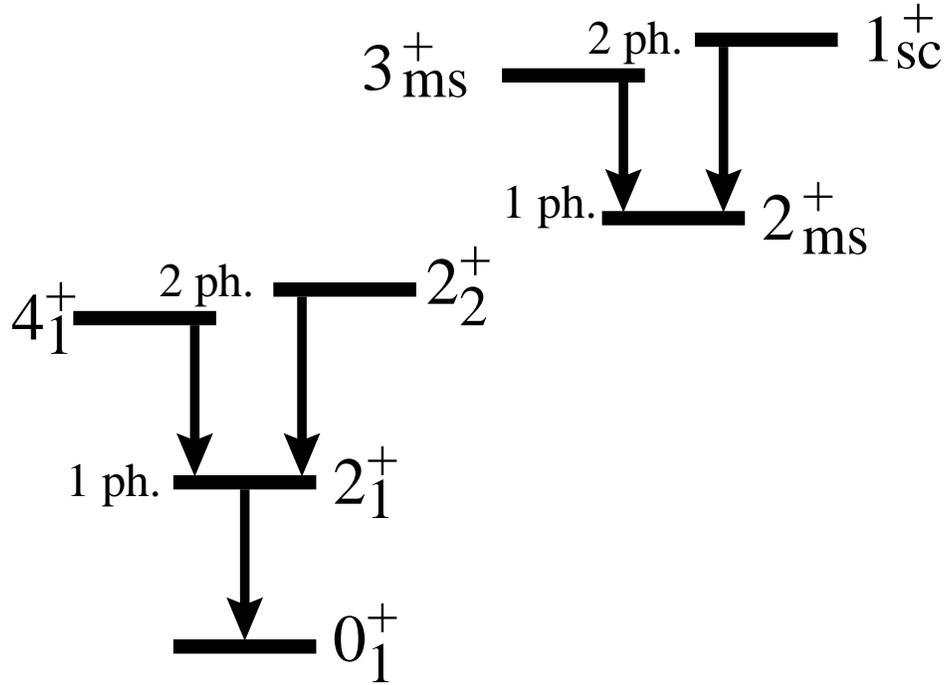}}
\caption{Sketch of the quadrupole vibrator scheme for 
         low-spin states of \protect$^{94}$Mo. 
         The arrows indicate transitions which are expected from 
         this scheme 
         to be collective \protect$E2$ transitions with large 
         isoscalar parts of the matrix elements. 
         The five corresponding values for the isoscalar $E2$ 
         matrix elements which were calculated in the shell model 
         are underlined in Table \protect\ref{transe2}.}
\label{fig:E2IS} 
\end{figure}

Having found support for MS assignments from the large isovector $M1$ 
transition strengths it is interesting to turn to the $E2$ transition 
properties. 
While the existence of large isovector $M1$ transitions may indicate a  
different proton-neutron symmetry, one can, in contrast, judge a 
similar proton-neutron symmetry for two states from the existence of 
collective 
isoscalar $E2$ transition matrix elements between the two states. 
The $E2$ transition operator is the sum of proton and neutron parts:
\begin{equation}
{\bf T}(E2)=e_p{\bf T}_p(E2)+e_n{\bf T}_n(E2), 
\end{equation} 
where $e_p$ and $e_n$ are the proton and the neutron effective quadrupole 
charges and 
${\bf T}_\rho(E2)=\sum_i(r_i^\rho)^2{\bf Y}_2(\theta^\rho_i,\phi^\rho_i)$. 
It is again convenient to decompose the $E2$ operator in an 
isoscalar part 
\begin{equation}
\label{is}
{\bf T}_{IS}(E2)={e_p+e_n \over 2} 
    \left[ {\bf T}_p(E2)+{\bf T}_n(E2) \right], 
\end{equation}     
and in an isovector part 
\begin{equation}
\label{iv}
{\bf T}_{IV}(E2)={e_p-e_n \over 2} 
    \left[{\bf T}_p(E2)-{\bf T}_n(E2) \right] \ . 
\end{equation} 
The calculated $E2$ transition strengths and matrix elements are compared 
to experiment and to schematic IBM-2 estimates in Tab. \ref{transe2}. 
In the low-lying low-spin level scheme of $^{94}$Mo one expects 
from a simple quadrupole vibrator picture the 
existence of the collective isoscalar $E2$ transitions 
which are indicated in Fig. \ref{fig:E2IS}.  
The transition from the $2^+_1$ state to the $0^+_1$ ground state is 
a collective isoscalar E2 transition. 
Many components of the wave function of the $2^+_1$ state contribute 
coherently to the matrix element of the $E2$ transition operator. 
The isoscalar part of the $\langle 2^+_1\parallel E2 \parallel 0^+_1\rangle$ 
matrix element is about ten times larger than the isovector part 
(see most right columns of Table \ref{transe2}). 
This is a well known general property of the lowest collective 
$2^+$ state. 
For $^{94}$Mo the $2^+_1$ state is calculated to exhaust  
97\% of the total isoscalar $E2$ excitation strength of the 
ground state to the $2^+$ states up to 4 MeV.  
The $4^+_1 \to 2^+_1$ and $2^+_2 \to 2^+_1$ transitions are of 
similarly collective dominantly isoscalar $E2$ character. 
The strong collective isoscalar $E2$ transitions between the $0^+_1$, $2^+_1$, 
$4^+_1$ and $2^+_2$ states prove the  proton-neutron symmetry of these states.

The E2 transitions from the $2^+_2$ state and the $2^+_3$ state to the ground 
state are much weaker than the $2^+_1 \to 0^+_1$ transition. 
The $2^+_2\to 0^+_1$ transition has isovector character but it is weak. 
The $2^+_3$ state, however, carries 10$\%$ of the total $E2$ 
excitation strength of the $2^+_1$ state and it is the largest 
$E2$ excitation above the $2^+_1$ state. 
This supports the one-phonon character of the $2^+_3$ state, 
which is suggested from its interpretation as the one-quadrupole phonon 
$2^+_{\rm ms}$ state. 
Indeed, the transition from the $2^+_3$ state to the ground state is 
a mixture of isoscalar and isovector parts with a notable isovector 
component.

The $E2$ matrix elements {\em between} various states with MS assignments are 
very interesting. 
The calculated wave functions of the $1^+_1$ state and the $3^+_2$ state are 
dominated by basis states with seniority $\nu=4$ supporting their 
two-phonon interpretation. 
The $1^+_1$ state shows no collective $E2$ transitions to the $2^+_1$, 
$2^+_2$ states. 
The $1^+_1\rightarrow 2^+_3$ transition is in contrast a collective isoscalar 
$E2$ transition which is comparable in strength with  the  
$2^+_1\rightarrow 0^+_1$ collective $E2$ transition. 
This indicates a similar proton-neutron symmetry of 
the $1^+_1$ state and the $2^+_3$ state and justifies to consider 
the $1^+_1$ state a two-phonon state formed by an isoscalar quadrupole 
phonon built on top of the MS $2^+_3$ state. 
The $2^+_3$ state has a proton-neutron symmetry similar to the $1^+_1$ state, 
but it is lower in energy and is of seniority $\nu=2$ like the collective 
lowest $2^+_1$ state. 
Also the calculated $3^+_2$ state decays by a strong collective isoscalar 
$E2$ transition to the $2^+_3$ state. 
The calculated $3^+_2\rightarrow 2^+_1$ $E2$ transition strength is five 
times smaller than the $3^+_2\rightarrow 2^+_3$ transition strength and 
the isovector part of the $E2$ matrix element is larger than the 
isoscalar part. 
Based on this comparison we can conclude that the $3^+_2$ state 
has qualitatively a similar proton-neutron symmetry as the $1^+_1$ state 
and the $2^+_3$ states. 
This conclusion supports the statement that the $3^+_2$ state of $^{94}$Mo 
is one more representative of proton-neutron collective states with a 
``mixed-symmetry'' character as it was argued before in Ref. \cite{PiFr3+}. 
We note, that also the calculated $3^+_2 \to 2^+_2$ $E2$ transition has a 
large isoscalar part, which, however, overestimates the data. 
This disagreement may again be caused by a too large mixing of the calculated 
$3^+_1$ and $3^+_2$ states and is probably due to a too large symmetric 
three-phonon component in the calculated wave function of the $3^+_2$ 
state.

Of particular interest in this article is the proton-neutron structure of 
those states to which mixed-symmetry was previously assigned from 
the measurements of large $M1$ and $E2$ transition strengths \cite{Pietr,PiFr3+}. 
In order to simplify the discussion we consider now a schematic 
model for the analysis of the wave function of the $2^+_{\rm ms}$ state, 
which reflects the logic of the IBM-2. 
Let us assume that valence protons and neutrons couple separately to 
collective $J^\pi_\rho=2^+_p$ and $J^\pi_\rho=2^+_n$ configurations with 
seniorities $\nu=2$. 
In a two-level model one collective $2^+_{s}$ state is formed 
by symmetric linear combination of the $2^+_p$ proton and $2^+_n$ 
neutron configurations: 
$|2^+_{s}>=(|2^+_p>+|2^+_n>)/\sqrt{2}$. 
The orthogonal linear combination 
$|2^+_{ns}>=(|2^+_p>-|2^+_n>)/\sqrt{2}$ has also seniority 
$\nu=2$ and is the collective ``nonsymmetric''  counterpart of the 
$|2^+_{s}>$ state. 
Furthermore, the $2^+_p$ proton and $2^+_n$ neutron configurations 
can couple to collective $J^\pi = 1^+,3^+$ states with seniority $\nu=4$ 
and isovector character of decay to the $2^+_{s}$ state.
The last three states should correspond to the lowest $1^+, 2^+, 3^+$ MS 
states in the IBM-2. 
The possible existence of such states in $^{94}$Mo and in neighboring 
nuclei as a result of quadrupole surface vibrations in anti-phase and 
corresponding two-phonon excitations was discussed earlier in a 
geometrical approach by A. Faessler \cite{Faess66}.

In the realistic shell model calculation for $^{94}$Mo presented above 
MS states cannot so easily be identified from the wave functions. 
Isospin symmetry and seniority conservation are broken due to the 
interaction chosen and the single particle orbitals considered. 
The ground state contains 72\% components with seniority $\nu=0$. 
The $2^+_1$ and $2^+_3$ states contain 70 $\%$ and 73 $\%$  components of 
seniority $\nu=2$. 
The relatively large components with seniority $\nu=2$ point 
at their predominantly one-quadrupole phonon nature \cite{Kim96,Pietr}. 
In contrast, the wave function of the $2^+_2$ state contains a large 
component of seniority $\nu=4$ (50\% of the wave function) in agreement 
with its usual two-quadrupole phonon interpretation.

Let us now further analyze the calculated wave functions of the 
$2^+_1$ state and the $2^+_3$ state, which are considered to 
represent well the symmetric and the MS one-phonon $2^+$ states 
of the IBM-2. 
It is interesting that the components of the $2^+_{1,3}$ states 
with seniority $\nu =2$, $|2^+_1,\nu=2\rangle$ and 
$|2^+_3,\nu=2\rangle$, are approximately orthogonal. 
Their normalized scalar product 
$$ \frac{\langle2^+_1,\nu=2|2^+_3,\nu=2\rangle} 
        {\sqrt{\langle2^+_1,\nu=2|2^+_1,\nu=2\rangle 
               \times \langle2^+_3,\nu=2|2^+_3,\nu=2\rangle } }
      = - 0.07 
$$ 
is small. 
This fact is 
not a trivial consequence of the orthogonality of the $|2^+_3\rangle$
and $|2^+_1\rangle$ eigenstates, because their wave functions 
contain noticeable components with higher seniority. 
The seniority $\nu$=2 components resemble the schematic 
symmetric $|2^+_{s}\rangle$ and ``nonsymmetric'' $|2^+_{ns}\rangle$ 
states discussed above and can be used for further analysis. 
For this purpose it is interesting to decompose the seniority 
$\nu$=2 components of the $2^+_1$ and $2^+_3$ states into their 
proton and neutron parts. 
One obtains 
\begin{equation}
\label{first}
|2^+_1,\nu=2\rangle=0.61|2^+_{1,p}\rangle+0.80|2^+_{1,n}\rangle
\end{equation} 
and 
\begin{equation}
\label{third}
|2^+_3,\nu=2\rangle=0.56|2^+_{3,p}\rangle-0.82|2^+_{3,n}\rangle \ . 
\end{equation}  
These decompositions cannot be directly compared to the schematic 
two level model mentioned above, because the basis states 
$|2^+_{i,\rho}\rangle$ are not identical. 
It turns out, however, that the normalized proton $|2^+_{3,p}\rangle$ basis 
state is rather similar to the normalized $|2^+_{1,p}\rangle$ basis state 
with a positive overlap of 
$\langle2^+_{3,p}|2^+_{1,p}\rangle$=0.98. 
However, the overlapping of the  neutron components of the  states  
(\ref{first}) and (\ref{third}) is smaller. 
It amounts only to $\langle2^+_{3,n}|2^+_{1,n}\rangle$=0.63 
indicating considerable deviations from the pure IBM-2 picture. 
For a direct comparison between the structure of the $|2^+_i,\nu=2\rangle$ 
components it is more useful to express the components 
$|2^+_{3,\rho}\rangle$ in Eq. (\ref{third}) by a linear combination of 
one part which is parallel to the $|2^+_{1,\rho}\rangle$ components 
and an orthogonal rest term $|R\rangle$ 
with $\langle R | 2^+_{1,\rho}\rangle \equiv 0$.   
We obtain 
\begin{equation}
|2^+_3,\nu=2\rangle = \gamma  
       \left[ 0.72 |2^+_{1,p}\rangle - 0.68 |2^+_{1,n}\rangle \right] 
      + |R\rangle 
\end{equation} 
This result can be interpreted in the following way: 
The dominant seniority $\nu = 2$ component of the $2^+_3$ state 
contains a fraction of $\gamma^2 = 58\%$ of components that form 
the dominant seniority $\nu = 2$ component of the $2^+_1$ state. 
Moreover, this fraction is almost orthogonal to the seniority 
$\nu = 2$ component of the $2^+_1$ state, because the proton part 
and the neutron part contribute with a different sign while their 
amplitudes are almost equal. 
Based on this observation one can consider the $|2^+_3\rangle$ state as 
a good realization of the collective $|2^+_{\rm ms}\rangle$ MS state.

The schematic analysis given above helps to make some more general 
conclusions about the nuclear structure properties which can lead 
to the appearance of the ``mixed symmetry'' states in near-spherical 
nuclei. 
At first, we remind that the amplitudes of proton and neutron parts of 
symmetric and nonsymmetric states have to be approximately equal.
These proton (neutron) parts of the wave functions of the symmetric 
state and its  nonsymmetric counterpart must be rather similar, too. 
As it follows from our calculations 
this condition can be achieved if the configurational space for valence 
protons (neutrons) may be restricted to one high-j orbital 
(in our case proton $1g_{9/2}$) and one of the nearest orbitals with 
small single particle j quantum number (in our case proton $2p_{1/2}$). 
An ideal case is a single high-j orbital. 
Otherwise, if there are few neighboring neutron (proton) orbitals with 
the j value comparable to the single particle angular momentum of the 
selected leading neutron (proton) orbital and if the influence of these  
orbitals cannot be neglected, then the neutron (proton) parts of the 
symmetric and nonsymetric states can be rather different. 
We can observe it already in our case for the neutron parts of the 
$2^+_1$ state and the $2^+_3$ state: the overlapping 
$\langle2^+_{3,n}|2^+_{1,n}\rangle$ is only 0.63. 
Therefore it can be expected that with the  increase of the number 
of valence neutrons in the considered configurational space  the neutron 
parts will be stronger fragmented and the neutron overlapping can be 
significantly reduced destroying the picture. On the contrary the increase of 
the number of valence protons in  $1g_{9/2}$ orbital will keep the proton 
parts similar and the ``mixed symmetry'' $2^+$ state will probably survive.
We hope that the above observations will be useful for the search for 
further mixed-symmetry phenomena in the neighboring nuclei.

We stress, however, that for the quantitative analysis of the relative 
proton-neutron symmetry of wave functions it can be more useful 
to analyze size and isotensor character of electromagnetic 
transition matrix elements between the calculated states, as 
was shown above, because of the presence of non-collective states 
in the shell model configurational space. 
Considering the radical truncation of the shell model problem which 
led to the formulation of the IBM it is remarkable how far the IBM 
and the shell model agree on the properties of mixed-symmetry states of 
$^{94}$Mo.

\section{Conclusions}

To summarize, we have performed shell model calculations 
for the near-spherical nucleus $^{94}$Mo using the Surface Delta 
Interaction as the residual interaction. 
We calculated excitation energies of the low-spin positive-parity states 
and $M1$ and $E2$ transition strengths between them. 
In most cases the calculations agree well with the data. 
Calculated wave functions, $M1$, and $E2$ matrix elements support 
the previous mixed-symmetry assignments for the $1^+_1$ state, the 
$2^+_3$ state, and the $3^+_2$ state of $^{94}$Mo. 
In particular, we find collective isoscalar $E2$ transitions 
between these three states and strong isovector $M1$ transitions  
to low-lying symmetric states. 
The strongest $M1$ transition is found between the $4^+_2$ state 
and the $4^+_1$ state. 
This transition is outside of the scope of the $sd$-IBM-2 approach. 
These findings indicate the common proton-neutron symmetry of 
the $1^+_1, 2^+_3, 3^+_2$ states showing that they form a class 
of states that differ to the lowest-lying ones by their proton-neutron 
structure. The analysis of the wave functions indicates in which 
neighboring nuclei these states can be most probably found.

\section{Acknowledgments}

We thank  R.F. Casten, E. Caurier, A. Dewald, L. Esser,  A. Gelberg, 
J. Ginocchio and T. Otsuka for discussions. R.V.J. thanks 
the Universit\"at zu K\"oln for a Georg Simon Ohm guest professorship.
This work was supported by the {\em Deutsche Forschungsgemeinschaft} 
under Contract Nos. Br 799/9-1 and Pi 393/1-1 and one of us 
(N.P.) got partly support by the US DOE under Contract No. 
DE-FG02-91ER-40609.


\begin{table}
\caption{Parameters used for the shell model calculation of 
         \protect$^{94}$Mo: proton and neutron single particle energies 
         for the orbitals included in the configurational space and 
         the interaction parameters of the Surface Delta Interaction 
         as defined in Ref. \protect\cite{bru77}.}
\label{parameters}
\begin{center}
\begin{tabular}{c|ccccccccccc}
Parameter & $\epsilon_{g_{9/2}}^p$ &  $\epsilon_{p_{1/2}}^p$ 
          & $\epsilon_{d_{5/2}}^n$ & $\epsilon_{s_{1/2}}^n$ 
          & $\epsilon_{g_{7/2}}^n$ & $\epsilon_{d_{3/2}}^n$ 
          & $\epsilon_{h_{11/2}}^n$ 
          & A$_{T=1}^{pp}$ & A$_{T=1}^{nn}$ & A$_{T=1}^{pn}$ 
          & A$_{T=0}^{pn}$ \\
\hline
 Value [MeV] & 0.0 & -0.8 & 0.0 & 1.4 & 2.0 & 2.2 & 2.4 
             & 0.31 & 0.24 & 0.27 & 0.50 
\end{tabular}
\end{center}
\end{table}

\begin{table}
\caption{Calculated structure of some low-lying eigenstates of the considered 
         shell model Hamiltonian. 
         The contributions of some low-seniority basis states, 
         which represent for some low-lying states the main 
         components of the wave functions, are shown.}
\label{structure}
\begin{center}
\begin{tabular}{l|ccccccccccc}
 \multicolumn{12}{c}{Component contributions to  wave functions ($\%$)}  \\
\hline
\multicolumn{1}{c}{Component} & \multicolumn{11}{c}{State, $J^\pi_i$}    \\
\hline
 & $0^+_1$ 
 & $0^+_2$ & $1^+_1$ & $1^+_2$ & $2^+_1$ & $2^+_2$ & $2^+_3$ & $3^+_1$& $3^+_2$ & $4^+_1$& $4^+_2$ \\
\hline
$\pi(g_{9/2}^2)_0\times\nu(d_{5/2}^2)_J$ & 34 & 20 &  &  & 17 & 3 & 37 &  & & 20 & 23 \\  
$\pi(p_{1/2}^{-2}g_{9/2}^4)_0\times \nu(d_{5/2}^2)_J$ & 17 & 21 &  &  & 8 & 0 & 11 &  & & 10 & 12 \\
$\pi(g_{9/2}^2)_J\times\nu(d_{5/2}^2)_0$ & --\tablenote{see two lines above} & --$^{\rm a}$ &  &  & 12 & 23 & 7 &  & & 13 & 21 \\  
$\pi(p_{1/2}^{-2}g_{9/2}^4)_J\times \nu(d_{5/2}^2)_0$ & --$^{\rm a}$ & --$^{\rm a}$ &  &  & 6 & 1 & 7 &  & & 6 & 4 \\
  $\pi(g_{9/2}^2)_2\times\nu(d_{5/2}^2)_2$ & 4 & 4 & 25 & 1 & 1 & 11 & 1 & 1 & 21 & 5 & 1 \\  
$\pi(p_{1/2}^{-2}g_{9/2}^4)_2\times \nu(d_{5/2}^2)_2$ & 2 & 6 & 8 & 0 & 0 & 5 & 1 & 0 & 3 & 2 & 0 \\
  $\pi(g_{9/2}^2)_4\times\nu(d_{5/2}^2)_4$ & 1 & 1 & 17 & 0 & 0 & 1 & 0 & 0 & 3 & 0 & 1 \\
  $\pi(g_{9/2}^2)_0\times\nu(d_{5/2}d_{3/2})_J$ &  &  & 14 & 2 & 0 & 0 & 0 & 0 & 0 & 3 & 0 \\
  $\pi(g_{9/2}^2)_0\times\nu(d_{5/2}g_{7/2})_J$ &  &  & 3 & 52 & 0 & 0 & 0 & 0 & 0 & 1 & 0 \\
$\pi(g_{9/2}^2)_0\times\nu(d_{5/2}s_{1/2})_J$ &  &  &  &  & 7 & 8 & 0 & 26 & 6 & 0 & 0 \\
$\pi(p_{1/2}^{-2}g_{9/2}^4)_0\times \nu(d_{5/2}s_{1/2})_J$ &  &  &  &  & 4 & 5 & 2 & 12 & 3 & 0 & 0 \\ \hline
 Components of seniority $\nu=0$ & 71 & 59 &  &  &  &  &  &  &  & &  \\
 Components of seniority $\nu=2$ &  &  & 21 & 54 & 71 & 50 & 73 & 40 & 11 & 65 & 71 \\
\end{tabular}
\end{center}
\end{table}


\begin{table}
\caption{Experimental \protect\cite{Pietr,PiFr3+} and calculated 
         M1 transition rates between low-lying states of \protect$^{94}$Mo. 
         The \protect$B(M1)$ values are given in units of 
         \protect$\mu_N^2$. 
         Quenched spin g-factors \protect$g^s_\rho=0.57 
         g^{s,{\rm free}}_\rho $ 
         were used in the shell model. 
         Hence, all \protect$M1$ transition strengths shown here 
         are generated by the isovector part of the \protect$M1$ 
         transition operator. 
         Schematic \protect$B(M1)$ values calculated 
         previously \protect\cite{Pietr,PiFr3+} 
         in the O(6) dynamical symmetry limit of the IBM--2 are given in 
         the last column.}
\label{transm1}
\begin{center}
\begin{tabular}{c|ccc}
$J_i\rightarrow J_f $ & \multicolumn{3}{c}{B(M1;$J_i\rightarrow J_f $) 
          ($\mu^2_N$)} \\
\hline 
 & Experimental & Shell Model &  IBM--2 \\
\hline
 $1_1^+ \rightarrow 0_1^+ $ & 0.16(1)    & 0.26  & 0.16   \\
 $1_2^+ \rightarrow 0_1^+ $ &0.046(18)   & 0.008 & 0      \\
 $1_1^+ \rightarrow 2_1^+ $ & $0.007^{+6}_{-2}$& 0.002 &0 \\
 $1_1^+ \rightarrow 2_2^+ $ & 0.43(5)   & 0.46 & 0.36     \\
 $1_1^+ \rightarrow 2_3^+ $ & $<$0.05   & 0.08 & 0        \\
 $2_2^+ \rightarrow 2_1^+ $ & 0.06(2)   & 0.094   &  0    \\
 $2_3^+ \rightarrow 2_1^+ $ &  0.48(6)  &  0.51 & 0.3     \\ 
 $2_4^+ \rightarrow 2_1^+ $ & 0.07(2)    & 0.01  & 0      \\ 
 $2_5^+ \rightarrow 2_1^+ $ & 0.03(1)    & 0.14  & 0      \\
 $3_2^+ \rightarrow 2_1^+ $ & $0.010^{+0.012}_{-0.006}$  & 0.10  & 0 \\
 $3_2^+ \rightarrow 4_1^+ $ & $0.074^{+0.044}_{-0.019}$  & 0.058  & 0.13 \\
$3_2^+ \rightarrow 2_2^+ $ & $0.24^{+0.14}_{-0.07}$  & 0.09  & 0.18 \\
$3_2^+ \rightarrow 2_3^+ $ & $0.09^{+0.07}_{-0.03}$\tablenote{ The E2/M1 multipole mixing ratio $\delta$ for the $3_2^+ \rightarrow 2_3^+ $ transition was
not measured unambiguously. For comparison to the SM results we use the smaller
value $\delta$=0.34(25) \protect\cite{PiFr3+}}  & 0.001  & 0 \\
 $4_2^+ \rightarrow 4_1^+ $ & 0.8(2)\tablenote{
see Ref. \protect\cite{Fr4+}}  & 1.79  &  -   
\end{tabular} 
\end{center}
\end{table}


\begin{table}
\caption{ Experimental \protect\cite{Pietr,PiFr3+} and calculated 
          \protect$E2$ transition rates between some low-lying 
          states of \protect$^{94}$Mo. 
          The shell model effective neutron and proton charges 
          $e_n=1.0e$ and $e_p=2.32e$ were used. The harmonic oscillator 
          length is b=A$^{1/6}$.  
          The  schematic IBM--2 estimates \protect\cite{Pietr,PiFr3+} 
          are given in the fourth column. 
          In columns five and six the total shell model \protect$E2$ 
          matrix elements are decomposed into their isoscalar and 
          isovector parts according to 
          Eqs. (\protect\ref{is},\protect\ref{iv}). 
          The underlined values for isoscalar matrix elements represent 
          cases for which collective isoscalar \protect$E2$ transitions 
          are expected from a vibrator model or from the 
          \protect$\gamma$-soft dynamical symmetry limits of the 
          IBM.}
\label{transe2}
\begin{center}
\begin{tabular}{c|ccc|cc}
$J_i\rightarrow J_f $ & \multicolumn{3}{c}{B(E2;$J_i\rightarrow J_f $),[e$^2$fm$^4$]} & \multicolumn{2}{c}{$<J_i||T_\rho(E2)||J_f>$,[efm$^2$]} \\
\hline 
 & Experimental & Shell Model &  IBM--2 & $\langle T_{IS} \rangle $ & $\langle  T_{IV} \rangle$ \\
\hline
 $2_1^+ \rightarrow 0_1^+ $ & 406(16) & 420 & 467 & \underline{50.3} & -4.6 \\
 $2_2^+ \rightarrow 0_1^+ $ & 6(2)    & 11  &  0  & -0.2 & 7.6 \\
 $2_3^+ \rightarrow 0_1^+ $ & 46(6)   & 42  & 30  & -6.6 & -7.8 \\
 $2_4^+ \rightarrow 0_1^+ $ & 5(1)    & 7.3 & 0   & 2.0 & 4.1 \\
 $2_5^+ \rightarrow 0_1^+ $ & 17(2)   & 8.4 & 0   & 6.0 & 0.5 \\
 $4_1^+ \rightarrow 2_1^+ $ &670(100) & 444 & 592 & \underline{-69.7} & 6.6\\
 $2_2^+ \rightarrow 2_1^+ $ & 720(260)& 482 & 592 & \underline{49.1} & 0 \\
 $2_3^+ \rightarrow 2_1^+ $ &$<$150   &0.03 &  0  & 4.3 & -4.0 \\
 $1_1^+ \rightarrow 2_1^+ $ & 30(10)  & 13  & 49  & 1.5 & 4.8\\
 $1_1^+ \rightarrow 2_2^+ $ & --      & 1.4 & 0   & -3.7 & 1.7 \\
 $1_1^+ \rightarrow 2_3^+ $ &$<$690   & 228 & 556 & \underline{25.7} & 0.5 \\
 $3_2^+ \rightarrow 2_1^+ $ & 10$^{+25}_{-9}$ & 43.2 & 48.3 & 5.3 & 12.0 \\
 $3_2^+ \rightarrow 4_1^+ $ & $<17.8$ & 23 & 0    & -12.8 & 0.1 \\ 
$3_2^+ \rightarrow 2_2^+ $ & $<$101.6 & 170.2 & 0 & -39.3 & 4.8 \\
$3_2^+ \rightarrow 2_3^+ $ & 254$^{+305}_{-203}$ \tablenote{see footnote ``a'' of Table \ref{transm1}} & 198.2 & 582 & \underline{42.5} & -5.3 
\end{tabular}
\end{center}
\end{table}

\end{document}